\documentclass[nofootinbib,twocolumn,showpacs,showkeys]{revtex4}
\usepackage{graphicx,color}
\usepackage{amssymb}
\usepackage{amsmath}
\usepackage{bm}
\usepackage{lipsum}
\usepackage{latexsym}
\usepackage{dcolumn}
\usepackage{float}
\usepackage{hyperref}

\begin{document}

\title{Revealing the non-adiabatic nature of dark energy perturbations from galaxy clustering data}

\author{Hermano Velten}\email{velten@pq.cnpq.br}
\author{Raquel Fazolo}\email{raquel.fazolo@gmail.com}

\affiliation{Universidade Federal do Esp\'{\i}rito Santo (UFES), Av. Fernando Ferrari S/N, 29075-910, Vit\'oria, Brazil}

\begin{abstract}

We study structure formation using relativistic cosmological linear perturbation theory in the presence of intrinsic and relative (with respect to matter) non-adiabatic dark energy perturbations. For different dark energy models we assess the impact of non-adiabaticity on the matter growth promoting a comparison with growth rate data. The dark energy models studied lead to peculiar signatures of the (non)adiabatic nature of dark energy perturbations in the evolution of the $f \sigma_{8}(z)$ observable. We show that non-adiabatic DE models become close to be degenerated with respect to the $\Lambda$CDM model at first order in linear perturbations. This would avoid the identification of the non-adiabatic nature of dark energy using current available data. Therefore, such evidence indicates that new probes are necessary to reveal the non-adiabatic features in the dark energy sector.

\end{abstract}

\maketitle

\section{\label{sec:level1}Introduction}

Within the standard cosmological model a successful General Relativity (GR) based description of the expanding universe demands the inclusion of two misterious components, namely dark matter (DM) and dark energy ({\rm DE}). While the former acts favoring structure formation composing 5/6 of the total matter in the universe, the latter manifests mostly at late times propelling the accelerating the background expansion. It is a well established fact we live in such accelerated epoch but the reasons behind it remain unknown, i.e., what is the nature of dark energy phenomena. Although the investigation of extended versions of GR as the explanation of the acceleration has called a huge attention in the last years \cite{Joyce:2016vqv}, the conventional approach for cosmology is still based on the idea that GR is sourced by a total energy-momentum tensor endowed with a {\rm DE} fluid component.   

Interpreting DE as a fluid requires adopting an equation of state relating its pressure ($p_{de}$) and energy density ($\rho_{de}$). The different dark energy models available in the literature correspond to choices for the equation of state parameter $w_{de}=p_{de}/\rho_{de}$. For the expanding background dynamics only the time dependence $w_{de}(t)$ is relevant. The simplest case $w_{de}=-1$ provides the same background dynamics as the inclusion of a cosmological constant ($\Lambda$) in the Einstein's equation. However, the growth of matter perturbations is affected by the existence of dark energy both via background effects, i.e., the slower the expansion, the faster the matter clustering (and vice versa), and also via the nature of dark energy perturbations $\delta p_{de}$. 

If the pressure of a fluid is written as $p\equiv p(\rho, \mathcal{S})$, where $\mathcal{S}$ means entropy, then its perturbation reads $\delta p = (\partial p / \partial \rho)_{\mathcal{S}=const.} \delta \rho + (\partial p / \partial \mathcal{S})_{\rho=const.} \delta S$, from which one identifies the adiabatic speed of sound $c^2_{ad} = (\partial p / \partial \rho)_{\mathcal{S}=const.}$ and the intrinsic non-adiabatic (entropic) contribution. For example, viscous fluids are charactherized by the appearence of the latter terms \cite{Blas:2015tla,Velten:2013qna,barrow,Velten:2011bg,Velten:2013pra,Velten:2014xca,Velten:2015tya,Carames:2014cga,Velten:2013rra,Velten:2012uv,HipolitoRicaldi:2010mf,HipolitoRicaldi:2009je}. It is worth noting then that background data such that Supernovae is not able to characterize non-adiabatic features in dark energy models.

Non-adiabaticity is also a typical feature of multi-component systems. Whether or not such components do interact via energy or momentum exchange, the existence of relative perturbations is expected in an expanding universe \cite{Brown:2011dn} and its properties have been studied in the literature \cite{Malik:2004tf, Christopherson:2008ry, Koshelev:2010wz, Zimdahl:2017cpa}. In order to model such relative perturbations, for instance, having any two components $x$ and $y$ one can generically define the relative entropic perturbations following $S_{xy}\sim(\delta x / \dot{x} - dy / \dot{y})$. The latter proportionality becomes an equality via some specific background choice as, for example, the Hubble expansion $H=\dot{a}/a$, where $a$ is the cosmological scale factor and the dot means derivative with respect to the cosmic time.

Our aim in this work is to assess the impact of the non-adiabatic nature of DE perturbations (specially the features pointed out above) on structure formation. By calculating the linear growth $f= d ln\,\delta_m / d\,ln a$ and the variance $\sigma^{2}_{R}$ of the density field smoothed at $R=8 h^{-1}$ Mpc scales we build the quantity $f \sigma_8$ in a bias independent manner. This allows us to promote a safe comparison between the theoretical predictions and available data. Such quantity is also useful from the observational perspective since it can be obtained from weak lensing and Redshift space distortions (RSD) analysis. In particular, we investigate how different dark energy models (taking into account their entropic perturbations) impact the $f \sigma_8$ observable. 

Next section is devoted to introduce the background and perturbative cosmological dynamics for different dark energy models. Our strategy presented in section \ref{sec:perturbations} relies on calculating the matter density perturbation sourced by the total gravitational potential. In computing the latter the non-adiabaticity (also including the dark energy density perturbations effects) is present. Then, we promote a direct comparison with $f\sigma_8$ data in section \ref{sec:Results}. We conclude in the final section. 

\section{\label{sec:level1}Dynamics of Dark energy models}
\subsection{\label{sec:background} Background Dynamics of Dark energy models}
Since we are interested in the late time aspects of structure formation we neglect the radiation effects. We investigate the total matter growth subjected to the recent effects caused by a dark energy component. We adopt a flat-FLRW expansion in which the expansion rate is given by 
\begin{equation}
\frac{H^{2}(a)}{H^{2}_0}=\frac{\Omega_{\rm m0}}{a^{3}}+(1-\Omega_{\rm m0})\,\, e^{-\int da\frac{1+w_{\rm DE}}{a}},
\end{equation}
where we will fix for our reference model the parameters $\Omega_{\rm m0}=0.3$ and $H_0=70 km \,s^{-1} Mpc^{-1}$. 

The dynamics of the background dark energy density is governed by the choice of the equation of state parameter (EoS) $w_{\rm DE}$. We will study some of the most relevant alternatives found in the literature. They are listed below:

\subsection{Dark energy models}

By assuming a equation of state for dark energy $w_{\rm DE}$ we completely determine the background dynamics (and also, as shown bellow, the perturbative dynamics). Some relevant parameterizations found in the literature are:

\begin{itemize}
\item The constant EoS  
\begin{equation}
w_{\rm DE}=w_0 ;
\end{equation}
\item The Chevallier-Polarski-Linder (CPL) \cite{Chevallier:2000qy,Linder:2002et}
\begin{equation}
w_{\rm DE}(a)=w_{0}+w_{1}(1-a);
\end{equation}
\item The Wetterich-logarithmic one \cite{Wetterich:2004pv}
\begin{equation}
w_{\rm DE}(a)=\frac{w_{0}}{\left[1+w_{1}{\rm ln}(1/a)\right]^2}.
\end{equation}
\end{itemize}

\subsection{\label{sec:perturbations} Perturbative Dynamics of Dark energy models}
We develop now a set of equations for the matter density perturbation $\delta_{\rm m}=\delta \rho_{\rm m} / \rho_{\rm m}$ which will be subjected to the peculiar existence of dark energy intrinsic and relative non-adiabatic perturbations. 

Following refs. \cite{Malik:2002jb, Bartolo:2003ad, Copeland:2006wr, Dent:2008ek}, we introduce the line element including scalar perturbations which is written according to
\begin{align}
ds^{2}=-(1+2A)dt^{2}+2a\partial_{i}Bdx^{i}dt \nonumber \\ \qquad
+a^{2}[(1+2\psi)\delta_{ij}+2\partial_{ij}E]dx^{i}dx^{j},
\end{align}
where $A, B, \psi$ and $E$ represents the scalar metric perturbations. According to the Newtonian (or longitudinal) gauge choice, i.e., $E=B=0$, the effective metric perturbations are described by the so called Newtonian potentials $\Phi$ and $\Psi$:
\begin{equation}
A-\frac{d}{dt}[a^{2}(\dot{E}+B/a)]\rightarrow A \equiv \Phi,
\end{equation}
\begin{equation}
-\psi+a^{2}H(\dot{E}+B/a)\rightarrow -\psi \equiv \Psi.
\end{equation}

Let us treat the entire cosmic substratum (Cold Dark matter + Dark energy) by an effective total fluid with density $\rho$ and pressure $p$. Then, the components of the energy momentum tensor of such effective one-fluid can be written as 
\begin{align}
T^0_0=-(\rho+\delta\rho), \quad T^0_\alpha-(\rho+p)v_{,a}, \nonumber \\
T^\alpha_\beta=(p+\delta p)\delta^{\alpha}_{\beta}+\Pi^{\alpha}_{\beta}.
\end{align}
In the absence of anisotropic stresses ($\Pi^{\alpha}_{\beta} =0$) -adopted here- both Newtonian scalar potentials coincide $\Phi=\Psi$. We are then left with the components of the Einstein's equation
\begin{equation}
-\frac{\nabla^2}{a^{2}}\Phi+3H^{2}\Phi+3H\dot{\Phi}=-4\pi G\delta\rho,
\end{equation}
\begin{equation}
H\Phi+\dot{\Phi}=4\pi Ga(\rho+p)v,
\end{equation}
\begin{equation}
3\ddot{\Phi}+9H\dot{\Phi}+(6\dot{H}+6H^{2}+\frac{\nabla^2}{a^{2}})\Phi=4\pi G(\delta\rho+3\delta p).
\end{equation}

Where $v$ (the fluid velocity potential in Newtonian gauge) is interpreted as the velocity of the fluid with respect to the normal line observers i.e., the magnitude of the fluid velocity relative to the Newtonian space.

In order to obtain a full set of equations for the perturbative dynamics it is also required the covariant conservation of the energy momentum tensor, i.e., $T^{\alpha}_{\beta \, ; \alpha}=0$. The continuity equation reads
\begin{equation}
\delta\dot{\rho}+3H(\delta\rho+\delta p)=(\rho+p)(3\dot{\Phi}+\frac{\nabla^2}{a}v),
\end{equation}
while the momentum conservation becomes
\begin{equation}
\frac{[a^{4}(\rho+p)v]^{\cdot}}{a^{4}(\rho+p)}=\frac{1}{a}\Big{(}A+\frac{\delta p}{\rho+p}\Big{)}.
\end{equation}

%\begin{equation}
%ds^{2}=-(1+2\Phi)dt^{2}+a^{2}(1-2\Phi)d\vec{x}^{2}
%\end{equation}
Working now in the Fourier space ($\nabla^2 \rightarrow -k^2$) we are also able to combine the above equations into useful forms. For example, we can write the total density contrast $\Delta = \delta \rho / \rho$ in terms of the potential $\Phi$ such that
\begin{equation}
\Delta=-\Big{(}\frac{2k^{2}}{3a^{2}H^{2}}\Big{)}\Phi-2\Phi
-2\frac{\dot{\Phi}}{H}.
\label{DeltaPhi}\end{equation}
Another useful relation involving the definition $\Theta = (c^2 - w)\Delta$ which emerges from Einsteins equation is
\begin{equation}
\Theta=\frac{2}{3H^{2}}[\ddot{\Phi}+H(4+3w)\dot{\Phi}+w\frac{k^{2}}{a^{2}}\Phi].
\end{equation}
We can also write down an equation for the evolution of $\Delta$ in terms of the velocity potential 
\begin{equation}
\dot{\Delta}=-3H\Theta+3(1+w)\dot{\Phi}-(1+w)\frac{k^{2}}{a}v,
\label{cont}
\end{equation}
as well as a dynamical equation for $v$ 
\begin{align}
\dot{v}=-vH(1-3w)-\frac{\dot{w}}{1+w}v+\frac{1}{a}\Big{[}\Phi+\frac{w}{1+w}\Delta \nonumber \\ \qquad+\frac{\Theta}{1+w}\Big{]}.
\label{Euler}
\end{align}
Finally, the full equation for the total density contrast (where no assumptions as for instance the quasi-static approximation have been done) reads
\begin{align}
\ddot{\Delta}+\dot{\Delta}(2-3w)H+\frac{k^{2}}{a^2}w\Delta+\frac{k^{2}}{a^{2}}(1+w)\Phi \nonumber \\
=3(1+w)[\ddot{\Phi}+\dot{\Phi}(2-3w)H]+3\dot{w}\dot{\Phi}-3H\dot{\Theta}\nonumber \\
+\Theta\Big{[}-\frac{k^{2}}{a^{2}}+\frac{3H^{2}}{2}(1+9w)\Big{]}
\end{align}

Following \cite{Bartolo:2003ad} the introduction of the intrinsic entropic perturbation of the dark energy component reads
\begin{equation}
\Gamma(a)\equiv\frac{3H(1+w_{\rm DE})c^{2}_{\rm a,DE}}{1-c^{2}_{\rm a,DE}}\Big{(}\frac{\delta\rho_{\rm DE}}{\dot{\rho}_{\rm DE}}-\frac{\delta p_{\rm DE}}{\dot{p}_{\rm DE}}\Big{)}.
\label{Gammadef}
\end{equation}

Also, the relative entropic perturbation for the system pressureless matter $+$ dynamical dark energy becomes
\begin{equation}
S(a)\equiv\frac{3H(1+w_{\rm DE})\Omega_{\rm m}}{1+w}\Big{(}\frac{\delta\rho_{DE}}{\dot{\rho}_{\rm DE}}-\frac{\delta\rho_{\rm m}}{\dot{\rho}_{\rm m}}\Big{)}.
\label{Sdef}
\end{equation}

It is worth noting that by decomposing the total pressure perturbation such that $\delta p=\delta p_{nad}+c^{2}_{a}\delta\rho$, the above relations allows us to write the intrinsic non-adiabatic pressure perturbation of the cosmic medium as
\begin{equation}
\delta p_{\rm nad}=\Omega_{\rm DE}[(-c^{2}_{\rm a,DE})S+(1-c^{2}_{\rm a,DE})\Gamma]\rho.
\end{equation}

The above definitions do not imply necessarily that energy transfer between the two components in the systems is imposed. Each component will obey a separate energy balance equation (with no sources). The definition for $\Gamma$ (\ref{Gammadef}) reflects aspects of the internal physical structure of the fluid DE while $S$ (\ref{Sdef}) quantities the multi-fluid nature of the global system. The standard adiabatic cosmology is recovered by setting $S = \Gamma = 0$.

%\begin{align}
%\ddot{\delta}+\dot{\delta}(2-3w)H+\frac{k^{2}}{a}w\delta+\frac{k^{2}}{a^{2}}(1+w)\Phi \nonumber \\ \qquad %=3(1+w)[\ddot{\Phi}+\dot{\Phi}(2-3w)H],
%\end{align}

%\begin{equation}
%\ddot{\delta}+2H\dot{\delta}-4\pi G\rho\delta=0.
%\end{equation}

%\begin{equation}
%g''+\Big{(}\frac{5}{2}-\frac{3w(a)\Omega_{DDE}(a)}{2}\Big{)}g'+\frac{3}{2}(1-w(a))\Omega_{DDE}(a)=0
%\end{equation}

The total fluid has an intrinsic adiabatic speed of sound given by
\begin{equation}
c_{\rm a}^{2}=\frac{\dot{p}}{\dot{\rho}}=\frac{w}{1+w}\Big{[}(1+w_{\rm DE})-\frac{a}{3}\frac{w\prime_{\rm DE}}{w_{\rm DE}}\Big{]},
\end{equation}
while the intrinsic dark energy adiabatic speed of sound reads
\begin{equation}
c^2_{\rm a, DE} = \frac{\dot{p_{\rm DE}}}{\dot{\rho_{\rm DE}}}=w_{\rm DE}-\frac{w_{\rm DE}^{\prime} a }{3(1+w_{\rm DE})}.
\end{equation}

The full perturbative dynamics is assessed after solving the following set of coupled equations for $\Phi, S$ and $\Gamma$,

\begin{align}
a^{2}H^{2}\Phi^{\prime\prime}+\Big{[}5aH+a^{2}H^{\prime}+3aHc_{a}^{2}\Big{]}a^{2}H \Phi^{\prime} \nonumber \\
\qquad +c_{\rm a}^{2}k^{2}\Phi+\Big{[}3+2a\frac{H^{\prime}}{H}+3c_{\rm a}^{2}\Big{]}a^{2}H^{2}\Phi= \nonumber \\
\qquad \frac{3}{2}a^{2}H^{2}\Omega_{\rm DE}[-c_{\rm a,DE}^{2}S+(1-c_{\rm a,DE}^{2}\Gamma],
\label{Pot1}
\end{align}

\begin{align}
aS^{\prime}=\Big{(}3w_{\rm DE}-\frac{3\Omega_{m}c_{\rm a,DE}^{2}}{1+w}\Big{)}S \nonumber \\
\qquad +\frac{3\Omega_{\rm m}(1+c_{\rm a,DE}^{2})\Gamma}{1+w}+\frac{k^{2}}{a^{2}H^{2}}\frac{S+\Gamma}{3} \nonumber \\
\qquad+\frac{k^{4}}{a^{4}H^{4}}\Big{(}\frac{2}{9}\frac{1+w_{\rm DE}}{1+w}\Big{)}\Phi,
\label{Pot2}
\end{align}
\begin{align}
a \Gamma^{\prime}=-\frac{3}{2}(1+w)S+3\Big{(}w_{\rm DE}-\frac{1+w}{2}\Big{)}\Gamma \nonumber \\ 
\qquad +\frac{k^{2}}{a^{2}H^{2}}\Big{(}-(1+w_{\rm DE})\mathcal{R}-\frac{S+\Gamma}{3}\Big{)} \nonumber \\
\qquad +\frac{k^{4}}{a^{4}H^{4}}\Big{(}-\frac{2}{9}\frac{(1+w_{\rm DE})}{1+w}\Phi\Big{)},
\label{Pot3}
\end{align}
where we have defined the gauge-invariant comoving curvature perturbation 
\begin{equation}
\mathcal{R}=\Phi+\frac{2}{3(1+w)}\Big{[}\Phi+a\frac{d\Phi}{da}\Big{]}.
\end{equation}

In the next section we will promote a comparison between the adiabatic (AD) and the non-adiabatic (NAD) situations. We call adiabatic dark energy model the potential $\Phi$ obtained solving Eq. (\ref{Pot1}) with vanishing right hand side i.e., $\Gamma=S=0$. 

For the NAD model we solve the coupled set of Eqs. (\ref{Pot1}) - (\ref{Pot3}) where $\Phi$ is sourced by the functions $S$ and $\Gamma$. The $k^4$ scale dependence seen in Eqs. (\ref{Pot1}) - (\ref{Pot3}) represents a new feature introduced by the non-adiabatic effects which manifest the fact that sub-horizon modes are more sensitive to the study of such effects.

We have also defined the total equation of state of the cosmic medium $w(a)=\sum\Omega_{i}(a)w_{i}(a)$. Since we are considering a pressureless matter field only the dark energy pressure plays a role. Therefore
%\begin{equation}
%w(a)=\sum\Omega_{i}(a)w_{i}(a),
%\end{equation}
\begin{align*}
w(a)&=\Omega_{\rm DE}(a)w_{\rm DE}(a) \\ 
&=\Omega_{\rm DE0}\Big{[}\frac{H_{0}^{2}}{H^{2}(a)}e^{-3\int da\frac{1+w_{DE}}{a}}\Big{]}w_{DE}(a).
\end{align*}

%\begin{equation}
%\delta p_{i}=\hat{c}_{s,i}^{2}\delta\rho_{i}+3aH(1+w_{i})(\hat{c}_{s,i}^{2}-c_{a,i}^{2})\rho_{i}v_{i}
%\end{equation}

%\begin{equation}
%\hat{c}_{s,DE}^{2}=c_{a,DE}^{2}+\frac{(1-c_{a,DE}^{2})\Gamma}{\delta_{DE}+3aH(1+w_{DE})v_{DE}}
%\end{equation}

\section{\label{sec:Results} Confronting matter perturbations with RSD data}

We compare now the predictions of the perturbative dynamics obtained in the last section with available data. It is important now to realize that we provided above an effective description for a one component model in which the total matter density $\rho$ is the sum of a pressureless matter component and dark energy. Therefore, the potential $\Phi$ is sourced by both the matter and dark energy density perturbations. The dark energy perturbative features are actually most captured via $c^2_{\rm a}$ and $c^2_{\rm a,DE}$ term which influences $\Phi$.
We are now interested however in the matter growth $\delta_m$.
After numerically solving Eqs. (\ref{Pot1}), (\ref{Pot2}) and (\ref{Pot3}) for the potential $\Phi$ we can use its solution to calculate the evolution of $\delta_m$.

We write an equation for the evolution of matter density perturbation $\delta_m$ as a function of $\Phi$. Since there is no interaction term between $\rho_{\rm m}$ and $\rho_{\rm DE}$ we adapt Eqs. (\ref{cont}) and (\ref{Euler}) for pressureless matter i.e., $w\rightarrow w_m=0$ (and making $\Delta \rightarrow \delta_{\rm m}$). Hence,
\begin{equation}
a^2 \delta^{\prime\prime}_m+\left(\frac{a H^{\prime}}{H}+3\right)a\delta^{\prime}_m+\frac{k^2 H^2_0}{a^2 H^2(a)}\Phi=0.
\end{equation}
We then calculate the quantities
\begin{equation}
f = \frac{\rm d \,ln \delta_m}{\rm d ln \,a}
\end{equation}
and
\begin{equation}
\sigma_8 = \sigma_8(a=1)\frac{\delta_{\rm m}}{\delta_{\rm m}(a=1)}.
\end{equation}
It is necessary to fix a fiducial value for the today's variance of power spectrum. Since we have adopted $\Omega_{\rm m0}=0.3$ the results from {\it Planck TE + low P} suggests the value $\sigma_{8}(a=1)=0.8$ \cite{Ade:2015xua}.

The different figures shown bellow refer to different dark energy models. In all the figures for the evolution of $f \sigma_8(z)$ we use 18 $f\sigma_8$ data points forming the {\it Gold} compilation proposed in Ref. \cite{Nesseris:2017vor}. Also, all figures bring the $\Lambda$CDM curve (solid black) as our reference model.

We solve the set of different equation for $\Phi$ generating initial conditions at a redshift $z_i=3000$ which corresponds to the onset of the matter dominated epoch. With help of the CAMB code, we calculate the power spectrum at $z_i$ for a fiducial $\Lambda$CDM model (the same plotted in the solid black line in all figures). The amplitude of density fluctuations is mapped into a $\Phi_i$ value. We also set adiabatic initial conditions $S(z_i)=\Gamma(z_i)=0$ for all models tested here. Then, the emergence of non-adiabatic effects in dark energy models is purely due to late time effects i.e., related to the dark energy equation of state. It is worth noting that as pointed out in Refs. \cite{Bartolo:2003ad, Dent:2008ek} from the structure of the perturbed equations (\ref{Pot1})-(\ref{Pot3}) the adiabaticity imposed by the initial conditions $S(z_i)=\Gamma(z_i)=0$ is preserved for superhorizon scales $(k/ a H \ll 1)$. Therefore, the appearance of non-adiabatic effects occurs only for modes which are of astrophysical interest well inside the horizon. In order to deal with the $k$-dependence of the equations we adopt the standard way to proceed which is to fix $k= 0.1 h Mpc^{-1}$ since this scale remains linear until today keeping the validity of our equations. Indeed, we have developed in the last section a set of equations valid for linear modes only. 

With the left panels of Figs.(1,2,3) we show how dark energy models influences the matter growth via the $f\sigma_8 \,{\rm x}\, z $ plane if we treat them an adiabatic (AD) or non-adiabatic (NAD) component. These figures also bring in the right panels the information about the conjoined expansion as discussed in \cite{Linder:2016xer}. %with the data provided by Ref. \cite{Basilakos:2017rgc}.

The labels in the figures indicate the parameters used in each curve. The general convention for all figures is such that solid lines represent adiabatic models while dashed lines the non-adiabatic cases.

Rather then a quantitative analysis, where e.g., a minimum $\chi^2$ is computed and best fit parameters are obtained, because the available data is still quite disperse (current errorbars are still large) we are only allowed to infer qualitative differences between adiabatic and non-adiabatic dark energy models.

\begin{figure*}
\centering
\includegraphics[width=0.4\textwidth]{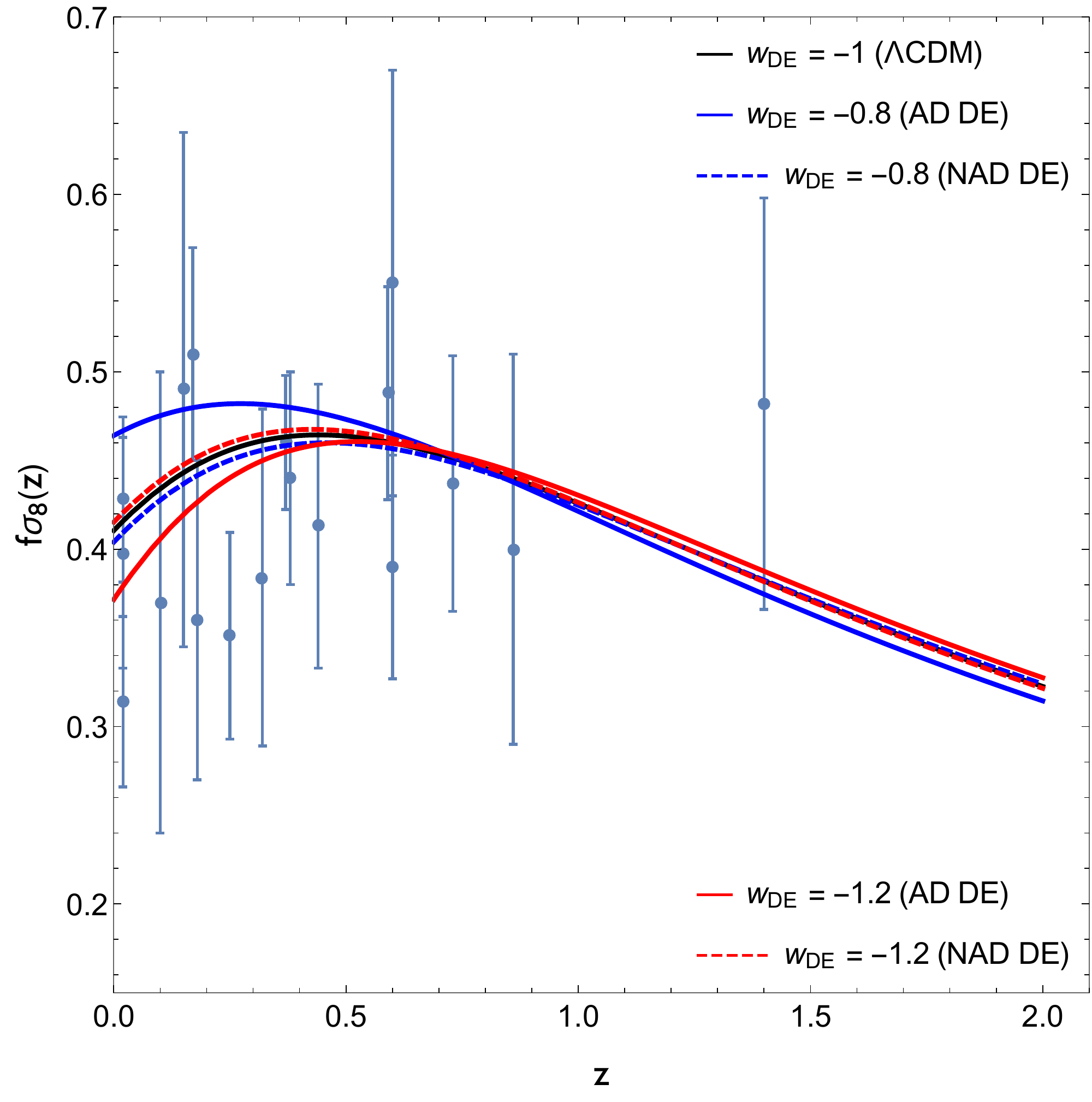}
\includegraphics[width=0.413\textwidth]{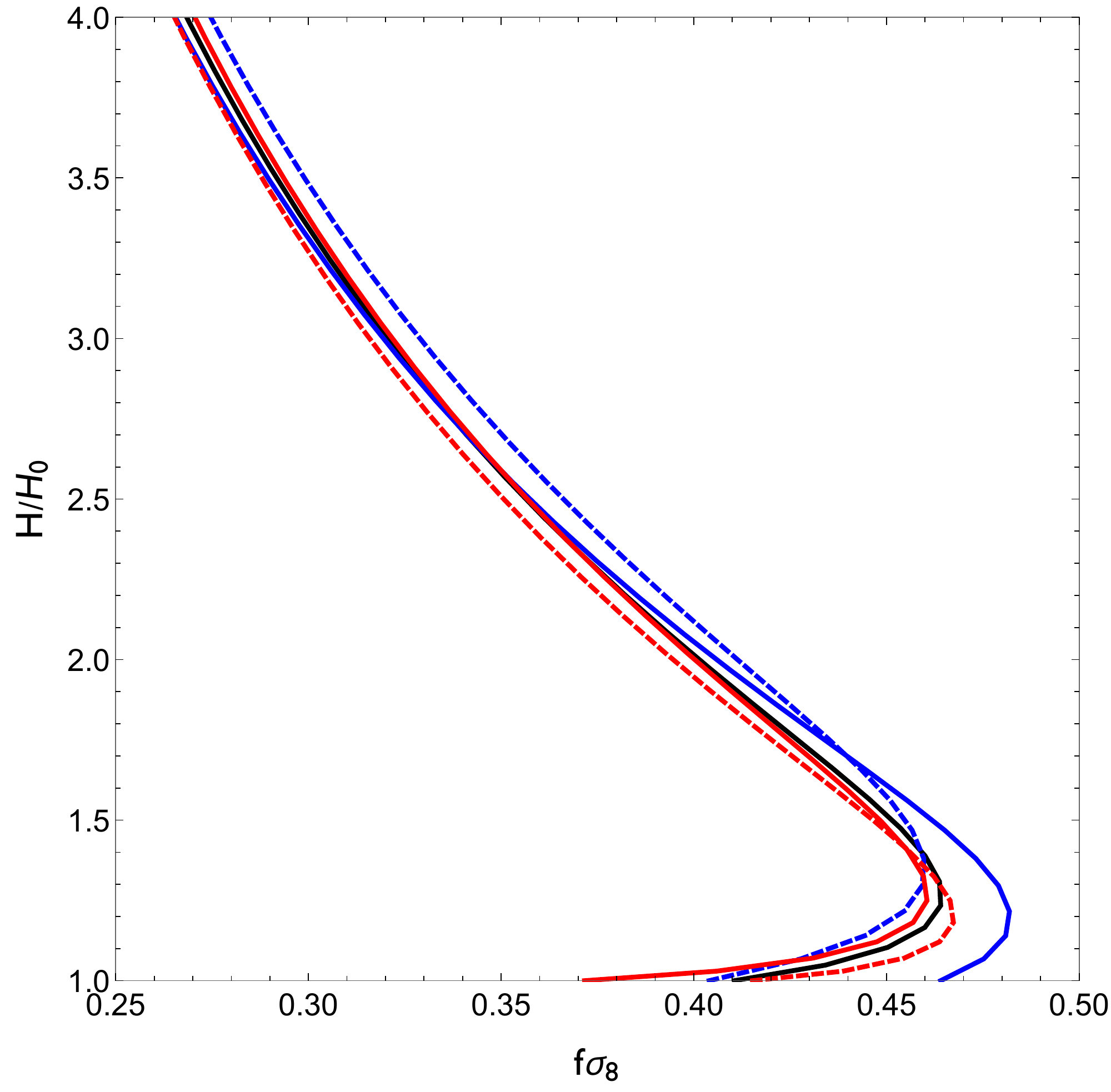}
\caption{Left: The matter clustering data $f\sigma_8$ as a function of the redshift. The 18 data points belong to the {\it Gold} sample (see text). Solid (dashed) lines computed with adiabatic (non-adiabatic) dark energy models. Right: The conjoined evolutionary tracks.}
\label{Figwconst}
\end{figure*}

In Fig. \ref{Figwconst} we compare the dark energy model with a constant equation of state with values $w_{DE}=-0.8$ (quintessence) and $w_{DE}=-1.2$ (phantom). Such values are extreme situations since beyond them (i.e., $w_{DE}>-0.8$ and $w_{DE}<-1.2$ ) there is no compatibility with other observational probes as for example Supernovae cosmology \cite{Betoule:2014frx}. As one can infer from (\ref{Gammadef}) and (\ref{Sdef}), the $\Lambda$CDM model ($w_{DE}=-1$) does not admit a non-adiabatic version since in this case $S=\Gamma=0$. But notice however that the same is no longer true for constant equations of state $w_{\rm DE} \neq -1$.
While the adiabatic version of these models (blue solid and red solid lines) presents a larger departure from the $\Lambda$CDM (black solid), the non-adiabatic counterparts tend to bring the curves closer to the standard cosmology. We tell in advance that this latter feature will also be seem in the remaining models. It is worth noting that the $\Lambda$CDM case seems to overestimate the magnitude of $f \sigma_8$ at low redshifts (see also \cite{Perenon:2015sla} for a discussion of this issue in a context of modified gravity). Then, the departure from standard $\Lambda$CDM model as given by the adiabatic phantomic case (solid red line) in Fig. \ref{Figwconst} seem to be preferred by the observed $f\sigma_8$ data. 

\begin{figure*}
\centering
\includegraphics[width=0.4\textwidth]{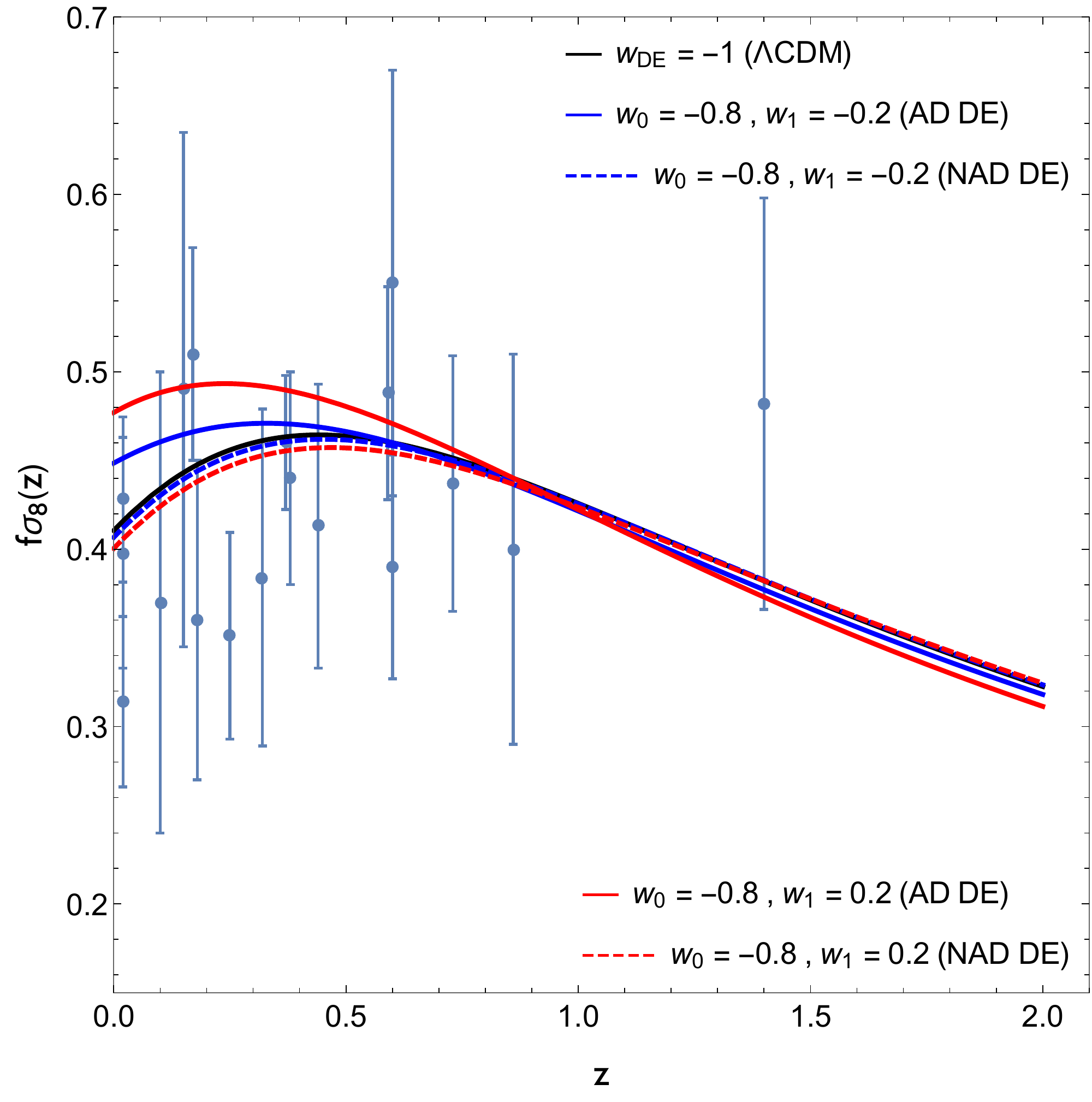}
\includegraphics[width=0.413\textwidth]{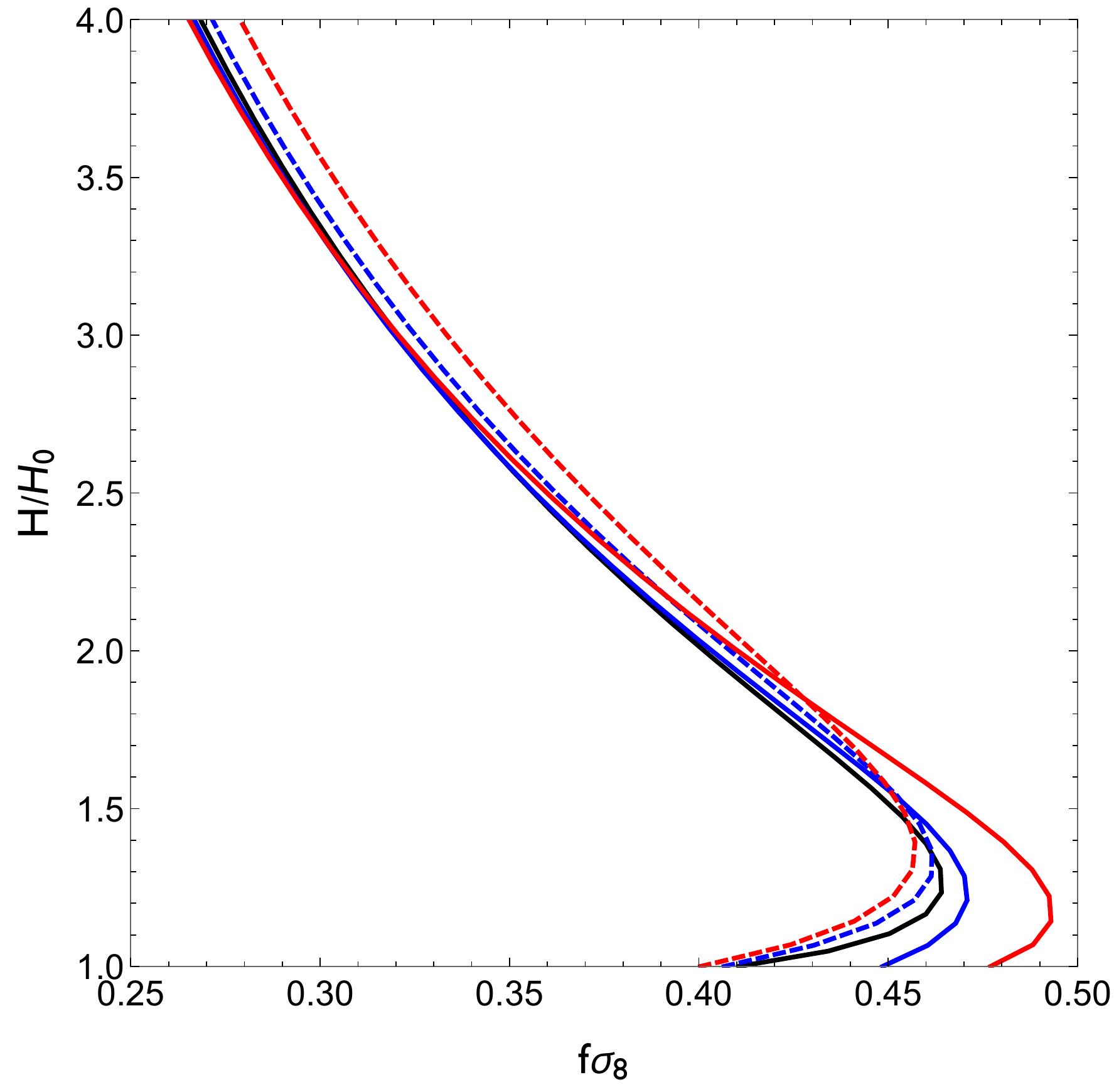}
\includegraphics[width=0.4\textwidth]{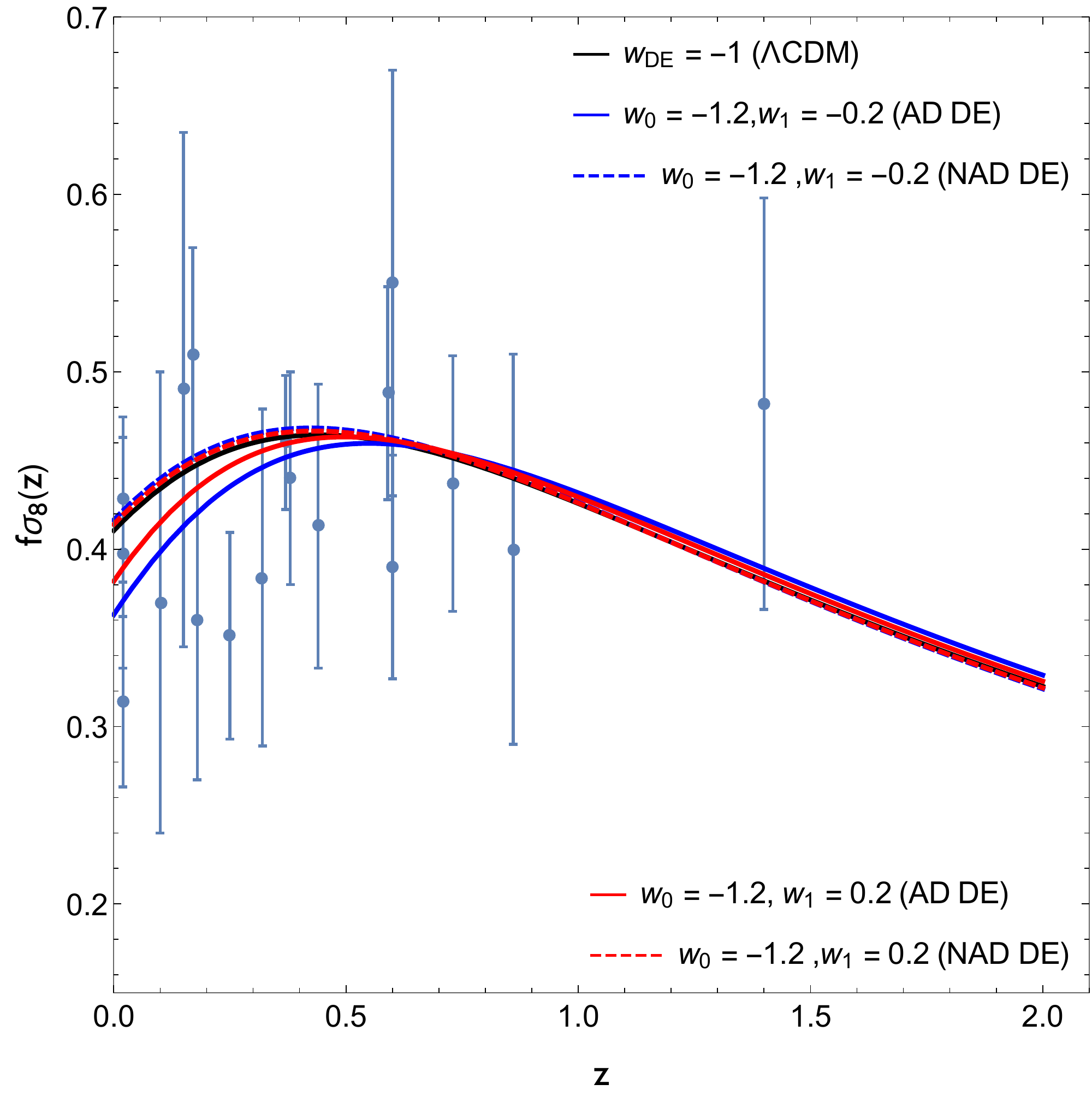}
\includegraphics[width=0.413\textwidth]{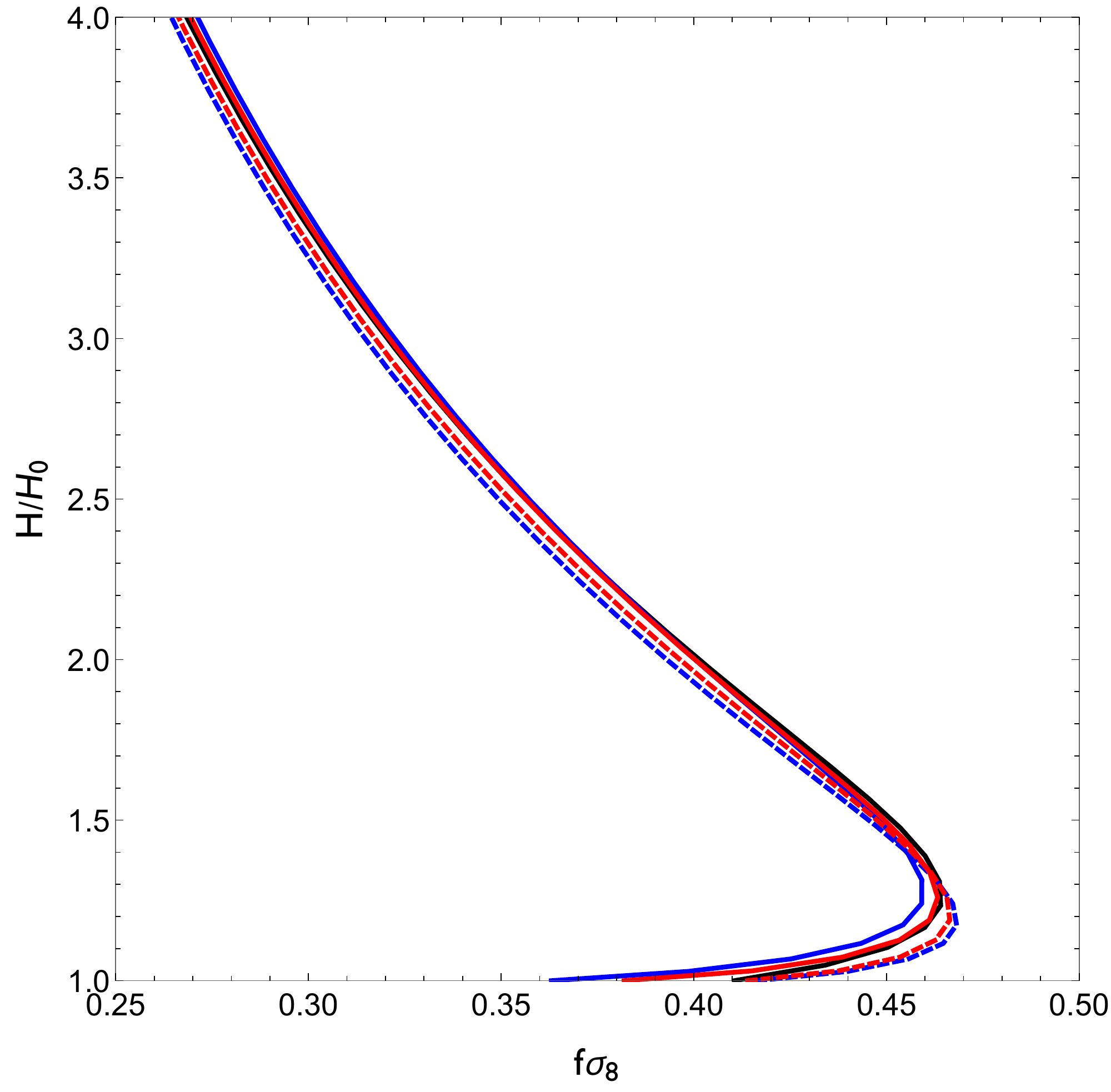}\caption{Left: The matter clustering (influenced by the CPL dark energy model) $f\sigma_8$ as a function of the redshift. The 18 data points belong to the {\it Gold} (see text). Solid (dashed) lines computed with adiabatic (non-adiabatic) dark energy models. Right: The conjoined evolutionary tracks.}
\label{FigwCPL}
\end{figure*}

\begin{figure*}
\centering
\includegraphics[width=0.4\textwidth]{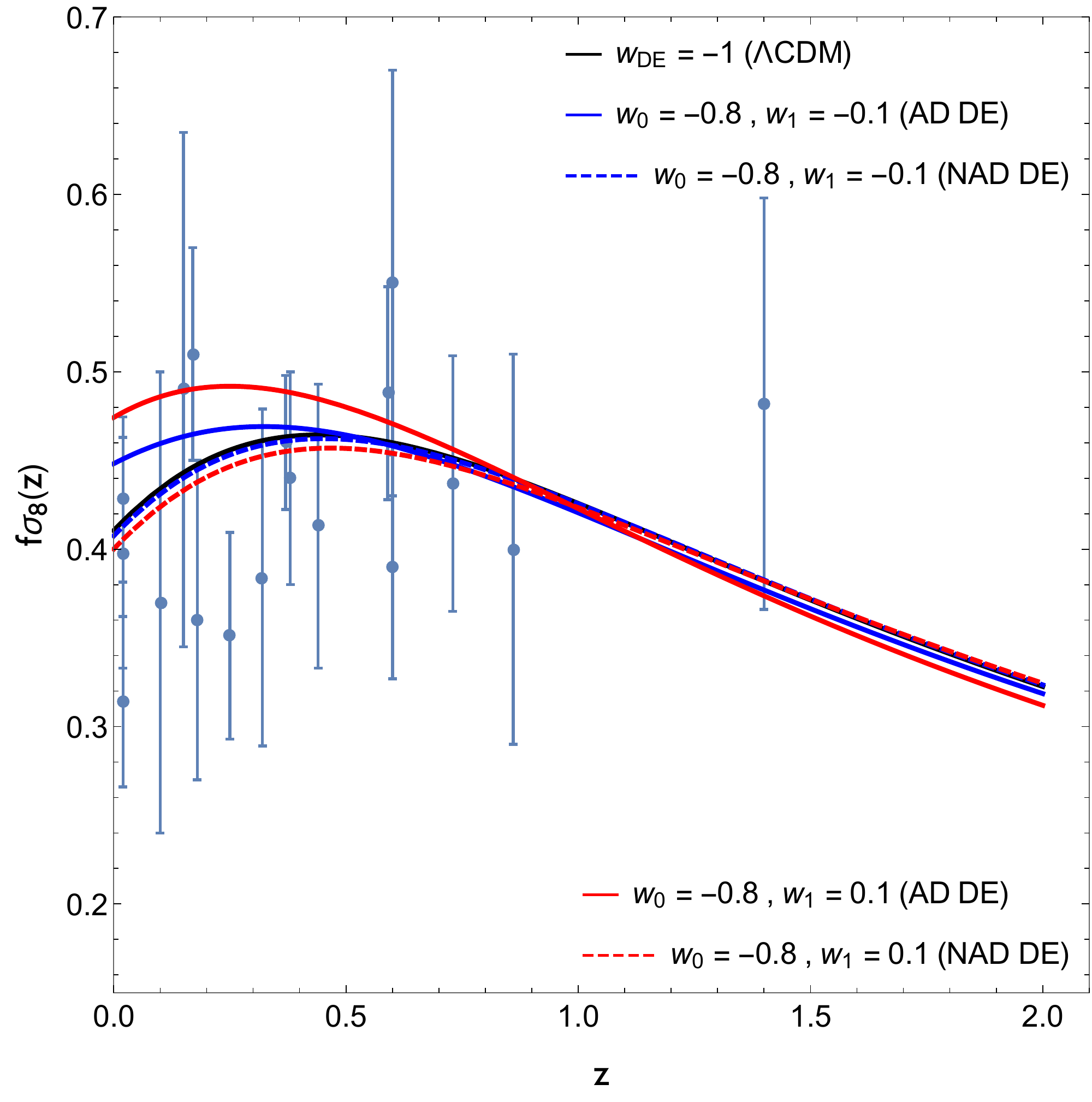}
\includegraphics[width=0.413\textwidth]{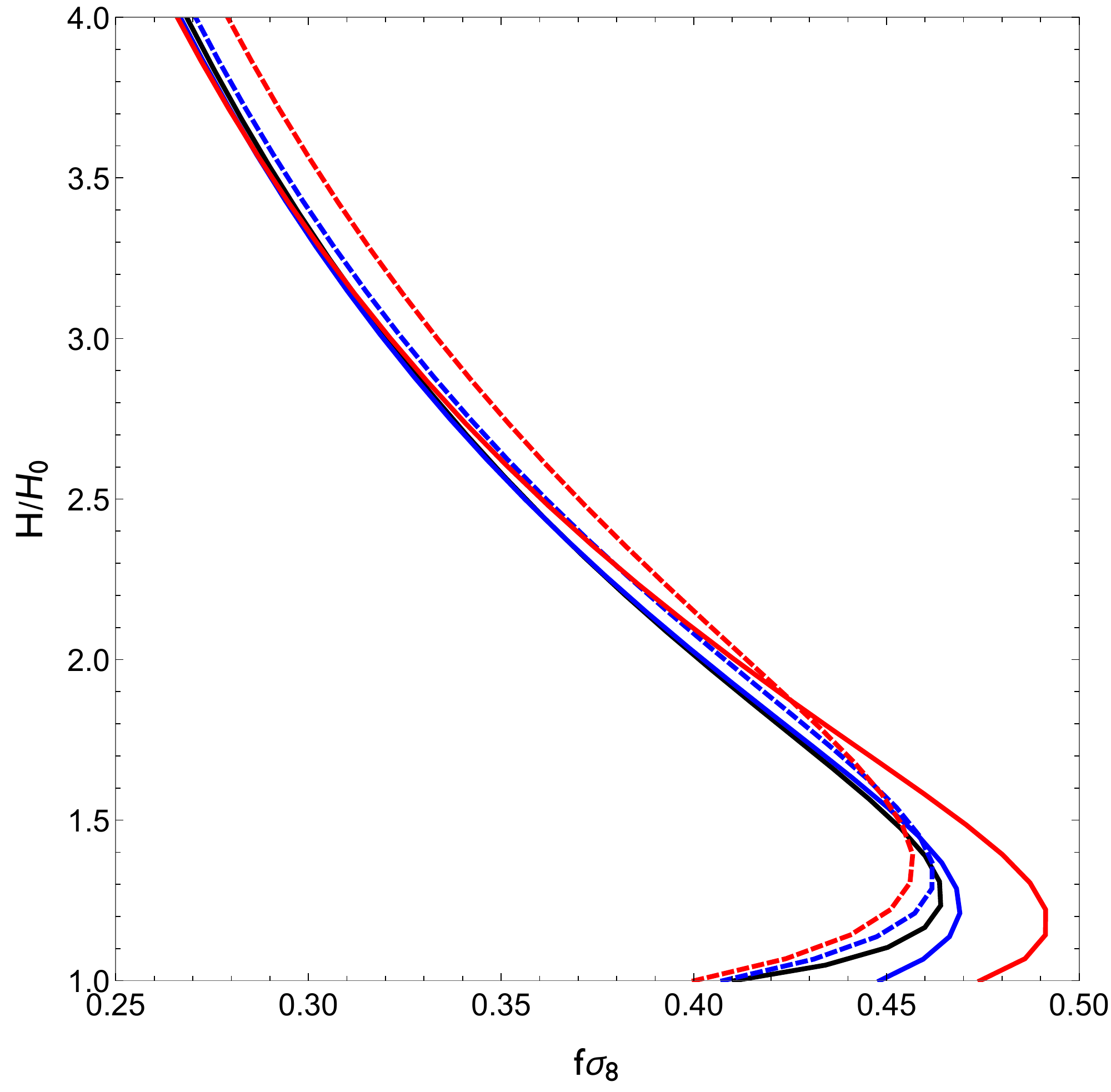}
\includegraphics[width=0.4\textwidth]{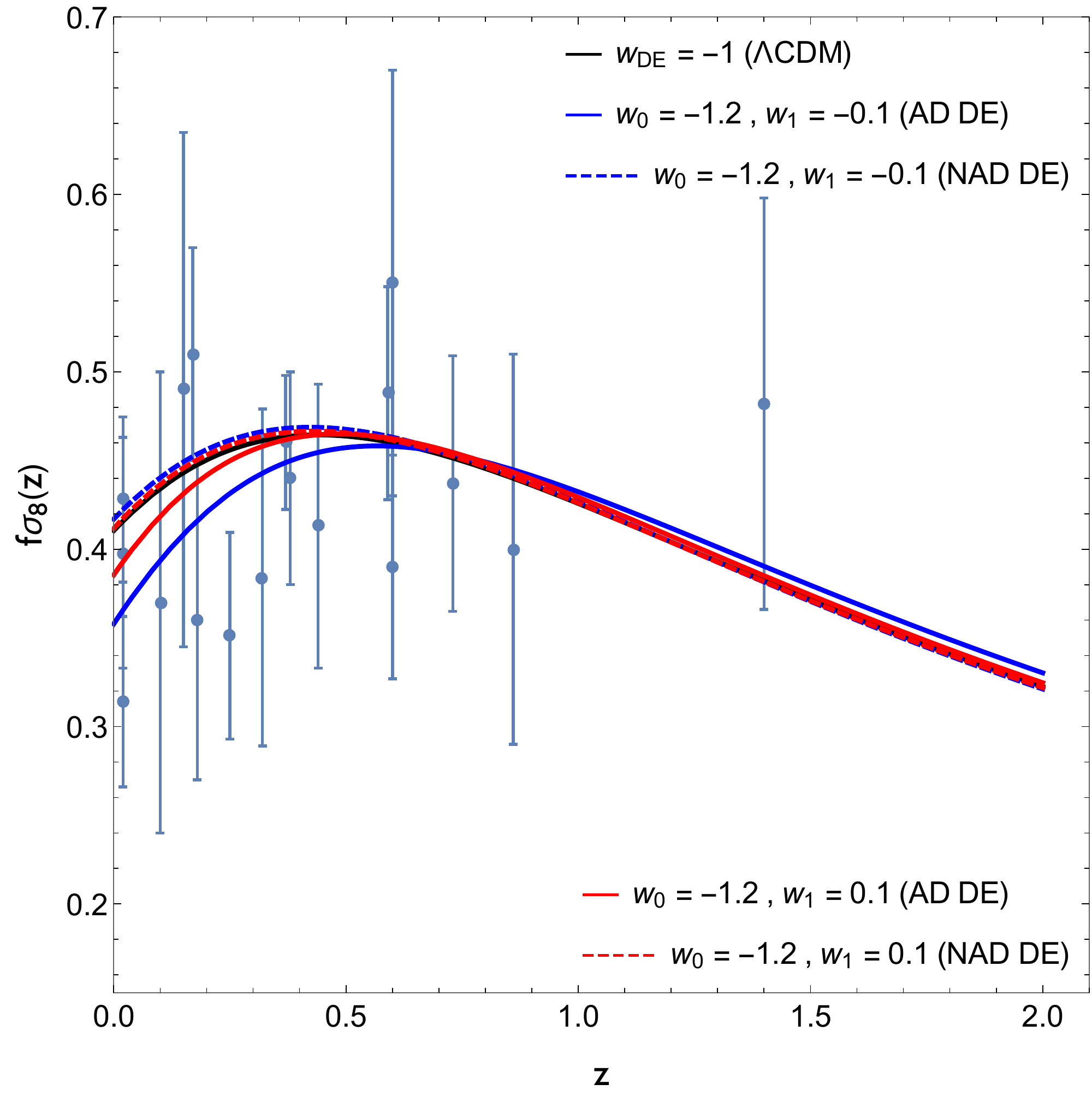}
\includegraphics[width=0.413\textwidth]{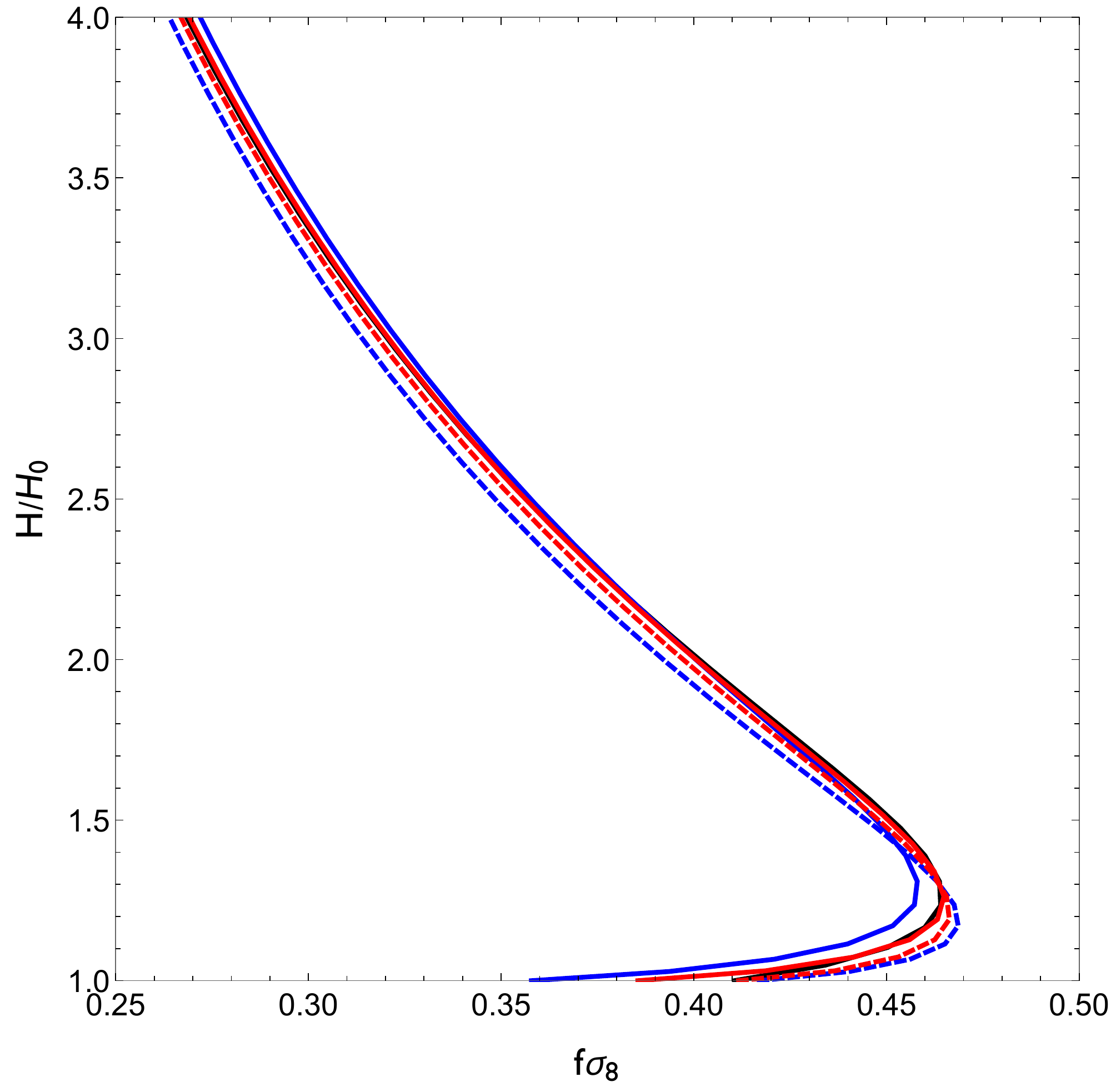}
\caption{Left: The matter clustering (influenced by the Wetterich-Logarithmic dark energy model) $f\sigma_8$ as a function of the redshift. The 18 data points belong to the {\it Gold} (see text). Solid (dashed) lines computed with adiabatic (non-adiabatic) dark energy models. Right: The conjoined evolutionary tracks.}
\label{FigwWP}
\end{figure*}

We study now time varying dark energy equations of state. The results for the CPL model are shown in Fig. \ref{FigwCPL} and for the Wetterich-logarithmic model in Fig. \ref{FigwWP}. All such models have fixed either $w_0=-0.8$ (upper panels) or $w_0=-1.2$ (bottom panels). The parameter $w_0$ corresponds to the todays's value of the dark energy equation of state. The time evolution is determined by $w_1$. We stress out how the non-adiabatic models remain close to the $\Lambda$CDM model.

\section{\label{sec:Conclusion}Conclusions}

In this work we have investigated whether the non-adiabatic nature of dark energy impacts the linear matter clustering. Although it is well know that on superhorizon scales $(k /aH <<1)$ initial adiabatic conditions remain unchanged, we have shown that for physical scales which are of astrophysical interest the non-adiabaticity features are able to emerge modifying the matter clustering patterns. We develop first order perturbative equations for matter overdensity ($\delta_m$) which are sourced by the gravitational potential $\Phi$. The usefulness of our approach relies in admitting that the existence of intrinsic and relative (with respect to matter) dark energy perturbation affects the evolution of potential $\Phi$. It is worth mentioning that this hypothesis is absent in the standard cosmology where only the adiabatic cases are considered. In light of the equations (\ref{Gammadef}) and (\ref{Sdef}) it becomes clear that the cosmological constant case $w_{\rm DE}=-1$ only presents a pure adiabatic behavior. Constant and time varying equations of state $w_{\rm DE} (a) \neq -1$ can however display a different behavior at the structure formation level if treated as adiabatic or non-adiabatic components.  

We have show that the matter growth data is particularly useful to distinguish and even to quantity the non-adiabaticity of dark energy models. The full distinguishability between adiabatic and non-adiabatic dark energy models is still limited since the errorbars in the current $f\sigma_8$ dataset samples is quite large. 

We have found however a common feature for all dark energy models investigated here. The matter growth for the non-adiabatic models tends to overlap with the $\Lambda$CDM prediction. Then, the use of clustering data would not be able to reveal specific signatures of dark energy non-adiabaticity. This means that dark energy can be actually a non-adiabatic time evolving component (with $w_0$ and $w_1$ parameters limited by the background tests) but its full thermodynamical nature is not distinguisable (since it is degenerated with $\Lambda$CDM) at linear level. We have also checked that the same happens for other parameterizations like the Jassal-Bagla-Padmanabhan \cite{JPB} and the Barboza-Alcaniz models \cite{BA}.

In order to shed some light on this issue the conjoined evolutionary tracks shown here can be useful but more accurate data is still needed.

The persistence of feature found in this work should be further explored with other cosmological probes. Also interesting is to assess the impact of non-adiabatic dark energy models in the full CMB spectrum and the cross-correlation galaxy-CMB where imprints of the Integrated Sachs-Wolfe effect appear. We leave this analysis for a future work.

%----------------------------------------------------------%
\section*{Acknowledgments}
%----------------------------------------------------------%
We acknowledge fruitful discussions with Winfried Zimdahl and Rodrigo von Maartens. We thank CNPq (Brazil) and FAPES (Brazil) for partial support.

\end{document}